\documentclass[aps,prl,twocolumn,showpacs,superscriptaddress]{revtex4}  
\usepackage{graphicx}  
\usepackage{dcolumn}   
\usepackage{bm}        
\usepackage{amssymb}   
\usepackage{epsfig}
\usepackage{slashed}

\usepackage{subfigure}

\begin{document}

\hspace{5.2in} \mbox{FERMILAB-PUB-11-486-E}

\title{$\boldsymbol{W\gamma}$ production and limits on anomalous $\boldsymbol{WW\gamma}$ couplings
in $\boldsymbol{p\bar{p}}$ collisions at $\boldsymbol{\sqrt{s} = 1.96}$~TeV }
%
\affiliation{Universidad de Buenos Aires, Buenos Aires, Argentina}
\affiliation{LAFEX, Centro Brasileiro de Pesquisas F{\'\i}sicas, Rio de Janeiro, Brazil}
\affiliation{Universidade do Estado do Rio de Janeiro, Rio de Janeiro, Brazil}
\affiliation{Universidade Federal do ABC, Santo Andr\'e, Brazil}
\affiliation{Instituto de F\'{\i}sica Te\'orica, Universidade Estadual Paulista, S\~ao Paulo, Brazil}
\affiliation{University of Science and Technology of China, Hefei, People's Republic of China}
\affiliation{Universidad de los Andes, Bogot\'{a}, Colombia}
\affiliation{Charles University, Faculty of Mathematics and Physics, Center for Particle Physics, Prague, Czech Republic}
\affiliation{Czech Technical University in Prague, Prague, Czech Republic}
\affiliation{Center for Particle Physics, Institute of Physics, Academy of Sciences of the Czech Republic, Prague, Czech Republic}
\affiliation{Universidad San Francisco de Quito, Quito, Ecuador}
\affiliation{LPC, Universit\'e Blaise Pascal, CNRS/IN2P3, Clermont, France}
\affiliation{LPSC, Universit\'e Joseph Fourier Grenoble 1, CNRS/IN2P3, Institut National Polytechnique de Grenoble, Grenoble, France}
\affiliation{CPPM, Aix-Marseille Universit\'e, CNRS/IN2P3, Marseille, France}
\affiliation{LAL, Universit\'e Paris-Sud, CNRS/IN2P3, Orsay, France}
\affiliation{LPNHE, Universit\'es Paris VI and VII, CNRS/IN2P3, Paris, France}
\affiliation{CEA, Irfu, SPP, Saclay, France}
\affiliation{IPHC, Universit\'e de Strasbourg, CNRS/IN2P3, Strasbourg, France}
\affiliation{IPNL, Universit\'e Lyon 1, CNRS/IN2P3, Villeurbanne, France and Universit\'e de Lyon, Lyon, France}
\affiliation{III. Physikalisches Institut A, RWTH Aachen University, Aachen, Germany}
\affiliation{Physikalisches Institut, Universit{\"a}t Freiburg, Freiburg, Germany}
\affiliation{II. Physikalisches Institut, Georg-August-Universit{\"a}t G\"ottingen, G\"ottingen, Germany}
\affiliation{Institut f{\"u}r Physik, Universit{\"a}t Mainz, Mainz, Germany}
\affiliation{Ludwig-Maximilians-Universit{\"a}t M{\"u}nchen, M{\"u}nchen, Germany}
\affiliation{Fachbereich Physik, Bergische Universit{\"a}t Wuppertal, Wuppertal, Germany}
\affiliation{Panjab University, Chandigarh, India}
\affiliation{Delhi University, Delhi, India}
\affiliation{Tata Institute of Fundamental Research, Mumbai, India}
\affiliation{University College Dublin, Dublin, Ireland}
\affiliation{Korea Detector Laboratory, Korea University, Seoul, Korea}
\affiliation{CINVESTAV, Mexico City, Mexico}
\affiliation{Nikhef, Science Park, Amsterdam, the Netherlands}
\affiliation{Radboud University Nijmegen, Nijmegen, the Netherlands and Nikhef, Science Park, Amsterdam, the Netherlands}
\affiliation{Joint Institute for Nuclear Research, Dubna, Russia}
\affiliation{Institute for Theoretical and Experimental Physics, Moscow, Russia}
\affiliation{Moscow State University, Moscow, Russia}
\affiliation{Institute for High Energy Physics, Protvino, Russia}
\affiliation{Petersburg Nuclear Physics Institute, St. Petersburg, Russia}
\affiliation{Instituci\'{o} Catalana de Recerca i Estudis Avan\c{c}ats (ICREA) and Institut de F\'{i}sica d'Altes Energies (IFAE), Barcelona, Spain}
\affiliation{Stockholm University, Stockholm and Uppsala University, Uppsala, Sweden}
\affiliation{Lancaster University, Lancaster LA1 4YB, United Kingdom}
\affiliation{Imperial College London, London SW7 2AZ, United Kingdom}
\affiliation{The University of Manchester, Manchester M13 9PL, United Kingdom}
\affiliation{University of Arizona, Tucson, Arizona 85721, USA}
\affiliation{University of California Riverside, Riverside, California 92521, USA}
\affiliation{Florida State University, Tallahassee, Florida 32306, USA}
\affiliation{Fermi National Accelerator Laboratory, Batavia, Illinois 60510, USA}
\affiliation{University of Illinois at Chicago, Chicago, Illinois 60607, USA}
\affiliation{Northern Illinois University, DeKalb, Illinois 60115, USA}
\affiliation{Northwestern University, Evanston, Illinois 60208, USA}
\affiliation{Indiana University, Bloomington, Indiana 47405, USA}
\affiliation{Purdue University Calumet, Hammond, Indiana 46323, USA}
\affiliation{University of Notre Dame, Notre Dame, Indiana 46556, USA}
\affiliation{Iowa State University, Ames, Iowa 50011, USA}
\affiliation{University of Kansas, Lawrence, Kansas 66045, USA}
\affiliation{Kansas State University, Manhattan, Kansas 66506, USA}
\affiliation{Louisiana Tech University, Ruston, Louisiana 71272, USA}
\affiliation{Boston University, Boston, Massachusetts 02215, USA}
\affiliation{Northeastern University, Boston, Massachusetts 02115, USA}
\affiliation{University of Michigan, Ann Arbor, Michigan 48109, USA}
\affiliation{Michigan State University, East Lansing, Michigan 48824, USA}
\affiliation{University of Mississippi, University, Mississippi 38677, USA}
\affiliation{University of Nebraska, Lincoln, Nebraska 68588, USA}
\affiliation{Rutgers University, Piscataway, New Jersey 08855, USA}
\affiliation{Princeton University, Princeton, New Jersey 08544, USA}
\affiliation{State University of New York, Buffalo, New York 14260, USA}
\affiliation{Columbia University, New York, New York 10027, USA}
\affiliation{University of Rochester, Rochester, New York 14627, USA}
\affiliation{State University of New York, Stony Brook, New York 11794, USA}
\affiliation{Brookhaven National Laboratory, Upton, New York 11973, USA}
\affiliation{Langston University, Langston, Oklahoma 73050, USA}
\affiliation{University of Oklahoma, Norman, Oklahoma 73019, USA}
\affiliation{Oklahoma State University, Stillwater, Oklahoma 74078, USA}
\affiliation{Brown University, Providence, Rhode Island 02912, USA}
\affiliation{University of Texas, Arlington, Texas 76019, USA}
\affiliation{Southern Methodist University, Dallas, Texas 75275, USA}
\affiliation{Rice University, Houston, Texas 77005, USA}
\affiliation{University of Virginia, Charlottesville, Virginia 22901, USA}
\affiliation{University of Washington, Seattle, Washington 98195, USA}
\author{V.M.~Abazov} \affiliation{Joint Institute for Nuclear Research, Dubna, Russia}
\author{B.~Abbott} \affiliation{University of Oklahoma, Norman, Oklahoma 73019, USA}
\author{B.S.~Acharya} \affiliation{Tata Institute of Fundamental Research, Mumbai, India}
\author{M.~Adams} \affiliation{University of Illinois at Chicago, Chicago, Illinois 60607, USA}
\author{T.~Adams} \affiliation{Florida State University, Tallahassee, Florida 32306, USA}
\author{G.D.~Alexeev} \affiliation{Joint Institute for Nuclear Research, Dubna, Russia}
\author{G.~Alkhazov} \affiliation{Petersburg Nuclear Physics Institute, St. Petersburg, Russia}
\author{A.~Alton$^{a}$} \affiliation{University of Michigan, Ann Arbor, Michigan 48109, USA}
\author{G.~Alverson} \affiliation{Northeastern University, Boston, Massachusetts 02115, USA}
\author{G.A.~Alves} \affiliation{LAFEX, Centro Brasileiro de Pesquisas F{\'\i}sicas, Rio de Janeiro, Brazil}
% ...
%
%
\collaboration{The D0 Collaboration\footnote{with visitors from
$^{a}$Augustana College, Sioux Falls, SD, USA,
$^{b}$The University of Liverpool, Liverpool, UK,
$^{c}$UPIITA-IPN, Mexico City, Mexico,
$^{d}$SLAC, Menlo Park, CA, USA,
$^{e}$University College London, London, UK,
$^{f}$Centro de Investigacion en Computacion - IPN, Mexico City, Mexico,
$^{g}$ECFM, Universidad Autonoma de Sinaloa, Culiac\'an, Mexico,
and 
$^{h}$Universit{\"a}t Bern, Bern, Switzerland.
$^{\ddag}$Deceased.
}} \noaffiliation
\vskip 0.25cm

\date{November 2, 2011}

\begin{abstract}
We measure the cross section and the difference in rapidities
between photons and charged leptons for inclusive 
$W(\to l\nu)+\gamma$ production in $e\gamma$ and $\mu\gamma$ final
states. Using data corresponding to an integrated luminosity of 4.2 fb$^{-1}$
collected with the D0 detector at the Fermilab Tevatron Collider,
the measured cross section times branching fraction for the process
$p\bar{p} \to W\gamma + X \to l\nu\gamma +X$ 
and the distribution
of the charge-signed photon-lepton rapidity difference are found
to be in agreement with the standard model. These results provide
the most stringent limits on anomalous $WW\gamma$ couplings for
data from hadron colliders: $-0.4 < \Delta\kappa_{\gamma} < 0.4$
and $-$0.08 $< \lambda_{\gamma} <$ 0.07 at the 95\% C.L.
\end{abstract}

\pacs{14.80.Bn, 13.85.Rm, 13.85.Qk}
\maketitle
The electroweak component of the standard model (SM) has been remarkably
successful in describing experimental results.
The $WW\gamma$ vertex is one example of self-interactions of
electroweak bosons that are a consequence of the
non-abelian $SU(2)_{L} \times U(1)_{Y}$ gauge symmetry of the SM.
In this Letter we use the process
$p \bar{p} \to W\gamma \to l\nu\gamma$ $(l = e, \mu)$ to
study this vertex and to search for any anomalous departure from 
SM $WW\gamma$ couplings.

An effective Lagrangian parameterizes the $WW\gamma$ couplings
with two parameters, $\kappa_{\gamma}$ and $\lambda_{\gamma}$ \cite{AC1, AC2},
under the assumptions of electromagnetic gauge invariance,
charge conjugation ($C$), parity ($P$) and $CP$ conservation.
The $\kappa_{\gamma}$ and $\lambda_{\gamma}$ couplings are related to the 
magnetic dipole and electric quadrupole moments 
of the $W$ boson \cite{AC1, AC2}. 
In the SM, $\kappa_{\gamma} = 1$ and $\lambda_{\gamma} = 0$,
and it is customary to introduce into the notation 
the difference $\Delta\kappa_{\gamma} \equiv \kappa_{\gamma} - 1$.

To assure that the $W\gamma$ cross section does not
violate unitarity, a form factor, with a common scale $\Lambda$ for each 
non-SM coupling parameter, is introduced to modify the terms as
$a_0 \to a_0/(1+\hat{s}/\Lambda^{2})^2$,
where $a_0 = \kappa_{\gamma}, \lambda_{\gamma}$, and
$\hat{s}$ is the square of the partonic center-of-mass energy. 
In this analysis, the scale $\Lambda$ is set to 2 TeV.
Contributions from anomalous couplings will increase the 
$W\gamma$ production cross section and yield photons of
higher energy than in the SM process\cite{AC2}.

\begin{figure}[htbp]
  \epsfig{file=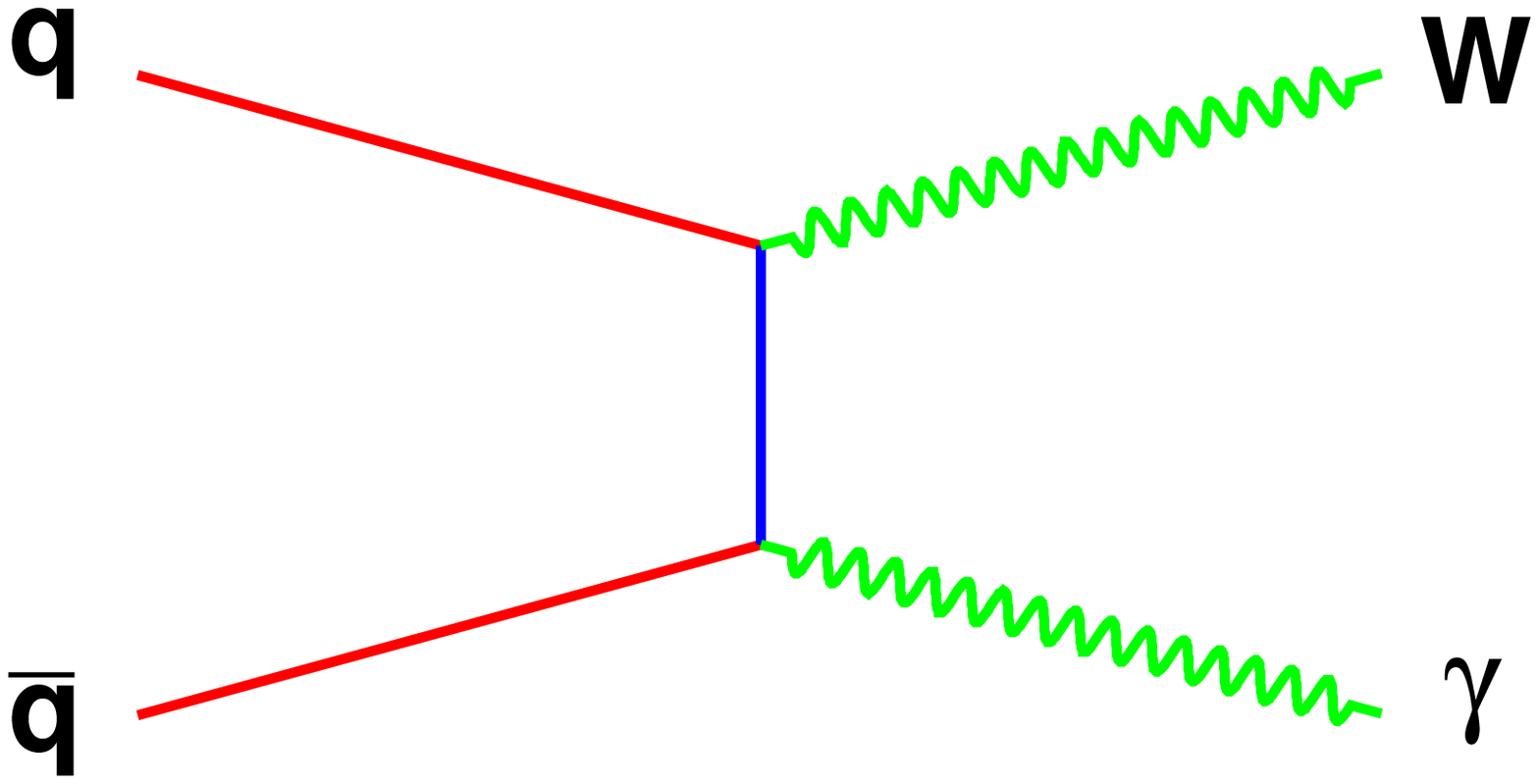,scale=0.2}
  \epsfig{file=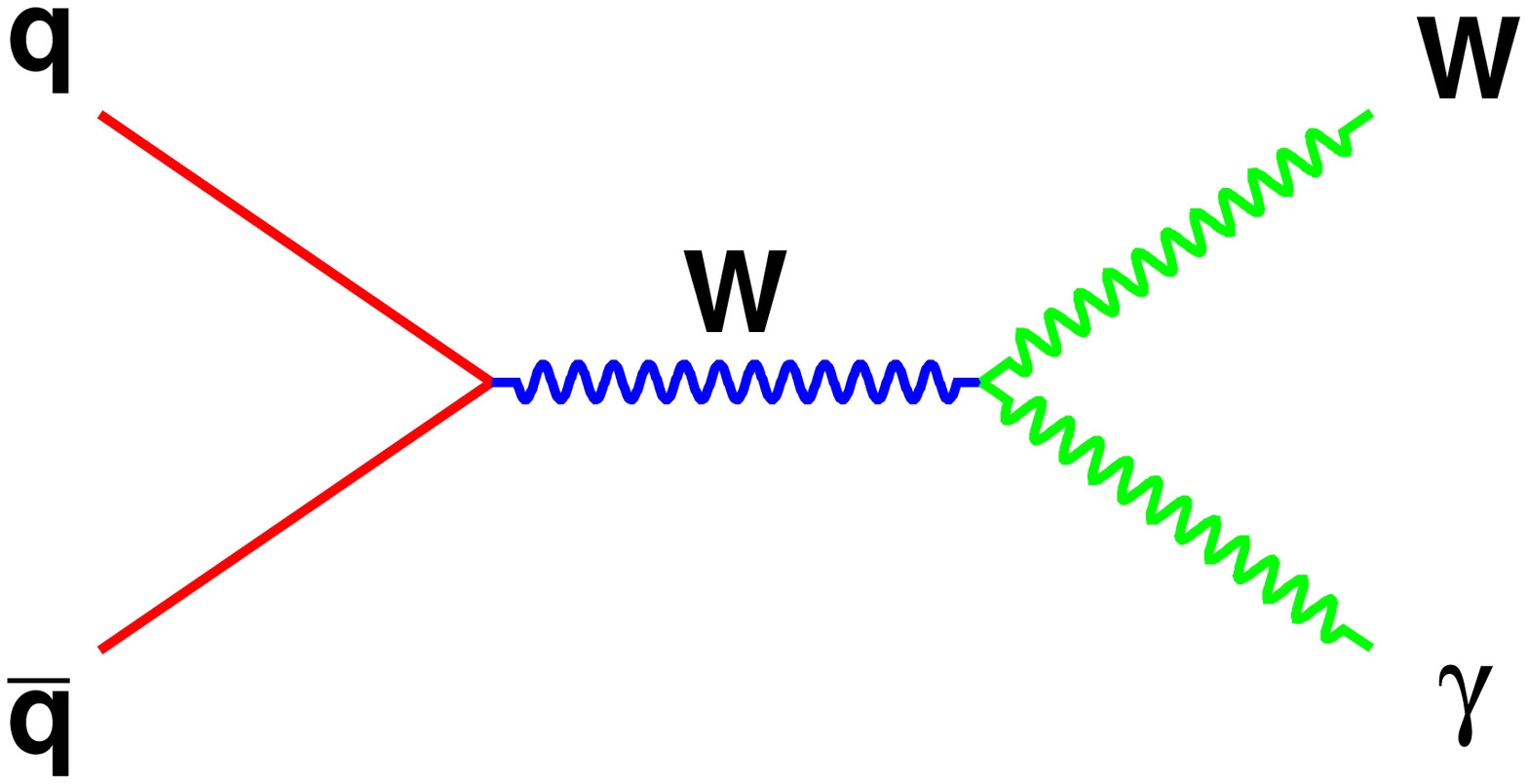,scale=0.2}
  \epsfig{file=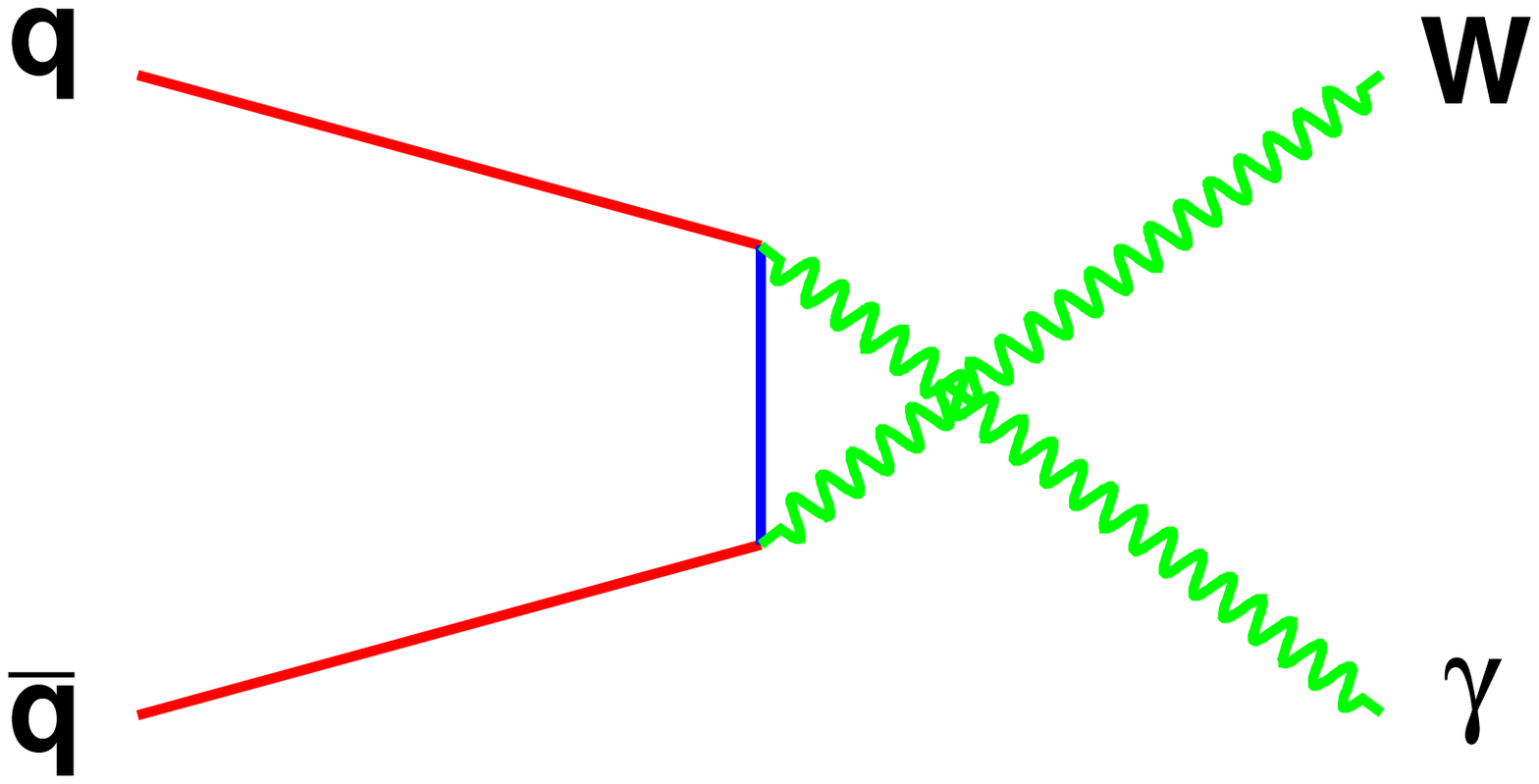,scale=0.2}
   \caption{\small (color online).
Feynman diagrams for prompt $W\gamma$ production.
}
  \label{figure-wg}
\end{figure}

In the SM, tree level production of a photon in association 
with a $W$ boson occurs due to 
prompt $W\gamma$ production via the diagrams shown in Fig.~\ref{figure-wg} or
via final state radiation (FSR), where a lepton from the $W$ boson decay
radiates a photon.
It is an important property of the SM prediction at leading order (LO) 
that the interference between the amplitudes 
in Fig.~\ref{figure-wg} produces a zero in the total $W\gamma$ yield
at a specific angle $\theta^{*}$ between the $W$ boson 
and the incoming quark \cite{raz-ref} in the $W\gamma$ rest frame.
Since in hadronic collisions the longitudinal momenta of neutrinos from $W$ decay
cannot be measured, the angle $\theta^{*}$ at which the radiation amplitude is zero
is difficult to measure directly.
However, the radiation amplitude zero (RAZ) is also visible in the charge-signed 
photon-lepton rapidity difference as a dip around $-$1/3 \cite{baur-ref2}.

In this Letter, we present measurements of the cross section
and the distribution of the charge-signed photon-lepton rapidity difference
for $W\gamma$ production as well as 
a search for anomalous $WW\gamma$ couplings,
using data corresponding to an integrated luminosity
of $4.2 \pm 0.3$ fb$^{-1}$
collected by the D0 detector at $\sqrt{s} = 1.96$ TeV 
at the Fermilab Tevatron Collider.
$W\gamma$ production has been studied previously 
at hadron colliders \cite{wg-UA2, wg-tev1, wg-tev2, wg-tev3, wg-cms}.
The latest published D0 result \cite{wg-tev3} 
represent the most stringent constraints
on anomalous $WW\gamma$ couplings, and include the first study
of the charge-signed photon-lepton rapidity difference at a hadron collider.
The results of the present analysis provide a significant improvement
in the sensitivity to $WW\gamma$ couplings through
a nearly factor of six increase in data and by
using an artificial neural network for photon identification.

The D0 detector \cite{d0det} comprises a central tracking
system in a 2~T superconducting solenoidal magnet,
surrounded by a central preshower (CPS) detector, a
liquid--argon sampling calorimeter, and an outer muon system.
The tracking system, a silicon microstrip tracker (SMT) 
and a scintillating fiber tracker (CFT), provides coverage 
for charged particles in the pseudorapidity range 
$|\eta| < 3$~\cite{d0_coordinate}.
The CPS is located immediately before the inner layer of the calorimeter,
and has about one radiation length of absorber, followed by several
layers of scintillating strips. 
The calorimeter consists of a central sector (CC) with
coverage of $|\eta|<1.1$, and two end calorimeters (EC)
covering up to $|\eta| \approx 4.2$. 
The electromagnetic (EM) section of the 
calorimeter is segmented into four longitudinal layers 
(EM$i$, $i=1,4$) with transverse
segmentation of 
$\Delta\eta\times\Delta\phi = 0.1\times 0.1$~\cite{d0_coordinate}, 
except in EM3, 
where it is $0.05\times 0.05$.
The muon system resides beyond the calorimeter and consists of a
layer of tracking detectors and scintillation trigger counters
before 1.8 T iron toroidal magnet, followed by two
similar layers after the toroid. The coverage of the muon
system corresponds to $|\eta| < 2$. 

Candidate events with the $W$ boson decaying into an electron and 
a neutrino are collected using a suite of single-electron triggers.
The electrons are selected by requiring an EM cluster in either
the CC ($|\eta| < 1.1$) or EC ($1.5 < |\eta| <2.5$) 
with transverse energy $E_T > 25$~GeV contained within 
a cone of radius 
$\Delta R=\sqrt{(\Delta \eta)^2 + (\Delta \phi)^2}=0.2$
centered on the axis of the EM shower. 
At least 90\% of the cluster energy must be deposited in the EM section
of the calorimeter.
In addition, electron candidates are required to
be isolated in the calorimeter and in the tracking detector, 
have a shower shape consistent with that
of an electron, and a spatial match to a track. 
A multivariate likelihood discriminant, which includes information
from the spatial track match, must be consistent with that for an electron.
An artificial neural network
is trained using information from the tracker, calorimeter, and  
CPS detectors to further reject background from jets misidentified as electrons.
The event missing transverse energy, $\slashed E_{T}$~\cite{MET-ref},
must exceed 25 GeV,
and the transverse mass of the $W$ boson, $M_T$~\cite{Mt-ref}, must exceed 50 GeV.
To reduce the background from $Z/\gamma^{*} \to ee$, where
an electron is misidentified as a photon because of tracking inefficiency,
the azimuthal angle between the electron and photon is required to be
$\Delta\phi_{e\gamma}<2$.

Candidate events with the $W$ boson decaying into a muon and a neutrino
are also collected using a suite of single-muon triggers.
The muons are required to be within $|\eta| < 1.6$, 
isolated in both the tracker and the calorimeter,
and matched to a track with transverse momentum $p_T > 20$ GeV.
To suppress the $Z/\gamma^{*} \to \mu\mu$ background, 
the $\slashed E_{T}$ in the event must exceed 20 GeV, 
$M_T$ must exceed 40 GeV, and there must be no additional muons 
or tracks with $p_T > 15$ GeV.
 
The photon candidates in both the electron and muon channels are 
required to have transverse energy $E_T^{\gamma} > 15$ GeV 
within a cone of radius $\Delta R=0.2$
centered on the EM shower. 
In addition, photon candidates are required to be either in the 
CC~($|\eta| < 1.1$) or EC~($1.5 < |\eta| <2.5$)
and must satisfy the following requirements:
(i) at least $90\%$ of the cluster energy is deposited in the
EM calorimeter; (ii) the calorimeter isolation variable
$I = [E_{\text{tot}}(0.4)-E_{\text{EM}}(0.2)]/E_{\text{EM}}(0.2) <$~0.15, where $E_{\text{tot}}(0.4)$ is the
total energy in a cone of radius $\Delta R=0.4$ and
$E_{\text{EM}}(0.2)$ is the EM energy in a cone of radius $\Delta R=0.2$;
(iii) the energy-weighted cluster width
in the EM3 layer is consistent with that for an EM shower;
(iv) the scalar sum of the $p_T$ of all tracks,
$p_{T_{\text{trk}}}^{\text{sum}}$, originating from the
interaction point in an annulus of $0.05<\Delta R<0.4$ 
around the cluster
is less than 2.0 (1.5) GeV for a cluster in the CC (EC);
(v) the EM cluster is not spatially matched to  
either a reconstructed track, or to energy depositions in the SMT or 
CFT detectors that lie along the trajectory
of an electron \cite{HOR}; 
and (vi) the output of an artificial neural network ($O_{NN}$)~\cite{gANN1}, 
that combines information from a set of variables sensitive to 
differences between photons and jets
in the tracking detector, the calorimeter, and the CPS detector, 
is larger than 0.75.
To suppress background from FSR, the photon and the lepton
must be separated by $\Delta R_{l\gamma} > 0.7$,
and the three-body transverse mass \cite{baur-ref2} of the photon, lepton, and 
missing transverse energy must exceed 110 GeV.
Furthermore, events are accepted only
if the $p\bar{p}$ interaction vertex is reconstructed within 60 cm of
the center of the D0 detector along the beam axis.

Trigger and lepton identification efficiencies are measured
using $Z \to ll$ $(l = e, \mu)$ data. Parts of the photon
identification efficiency which rely exclusively on 
calorimeter information are also determined using
$Z \to ee$ data.  The photon track veto efficiencies are 
determined using $Z \to l^{+}l^{-}\gamma$ events,
where the photons are radiated from charged leptons in $Z$ boson decays.
The selected data sample is contaminated by the following backgrounds: 
(i) $W+$jet events with a jet misidentified as a photon;
(ii) ``$leX$'' events with a lepton, electron, and $\slashed E_{T}$
with the electron misidentified as a photon; (iii)
$Z\gamma \to ll\gamma$ events containing 
an unidentified lepton; and (iv) $W\gamma \to \tau\nu\gamma$ events
with the $\tau$ decaying to $e$ or $\mu$.
A Monte Carlo (MC) simulation is used to model 
the $W\gamma$ signal, as well as the $Z\gamma \to ll\gamma$ and 
$W\gamma \to \tau\nu\gamma$ backgrounds.
The signal is simulated using the 
Baur and Berger LO event generator~\cite{AC2}, 
interfaced to {\sc pythia}~\cite{pythia-ref} 
for subsequent parton showering and hadronization.
The shape and normalization of the signal $E_T^{\gamma}$ spectrum
are reweighted to the next-to-leading order (NLO) prediction~\cite{baur-ref2}.
The acceptance of the kinematic and geometric requirements
for this analysis is calculated using 
this $E_T^{\gamma}$-weighted signal MC. 
The $Z\gamma \to ll\gamma$ and $W\gamma \to \tau\nu\gamma$ 
background processes are simulated with {\sc pythia}.
All MC events are generated using the CTEQ6L1 \cite{cteq-ref} 
parton distribution functions (PDF), 
followed by a {\sc geant} \cite{geant-ref} 
simulation of the D0 detector.
To accurately model the effects of multiple $p\bar{p}$ interactions
and detector noise, 
data from random $p\bar{p}$ crossings
are overlaid on the MC events. 
The instantaneous luminosity spectrum of these overlaid events 
is matched to that of the events used in the data analysis.
The simulated events are processed using the
same reconstruction code that is used for data.

$W+$jet production is the dominant background for both the electron and muon channels.
To estimate this background, the fraction of jets that pass the photon selection criteria but
fail either the $p_{T_{\text{trk}}}^{\text{sum}}$ or the shower width requirement,
as determined by using a multijet data sample, 
is parameterized as a function of $E_T^{\gamma}$ and $\eta_{\gamma}$.
The background from $W+$jet production is then estimated starting from an data sample 
obtained by reversing the requirements either on 
$p_{T_{\text{trk}}}^{\text{sum}}$ or on shower width requirement, and
applying the same parameterization.
As a cross-check, the $W+$jet background is also estimated through a fit 
to the $O_{NN}$ distribution in data,
using MC templates constructed from generated photon and jet events. The
result is in good agreement with that obtained from
the ratio method.
The ``$leX$'' background is also estimated from an orthogonal data sample by 
requiring the electron candidate
to be matched to a high-quality track. The number of ``$leX$'' events is obtained 
by using this orthogonal data sample, taking into account the ratio of the
track--matching inefficiency to the track matching efficiency 
obtained from $Z \to ee$ data.
Events in the electron channel that have
both the electron and photon in the EC are excluded from this analysis,
because of the poor acceptance for signal and the presence of overwhelming background.
The number of predicted and observed events in both the electron
and muon channels are summarized in 
Table \ref{table:Nevents}.

\begin{table}[t]
\begin{center}
\begin{tabular}{ccc}
    \hline \hline
    & $e\nu\gamma$ channel& $\mu\nu\gamma$ channel \\
    \hline
    $W+$jet& 33.9 $\pm$ 3.7 & 64.6 $\pm$ 6.8 \\
    $leX$ & 1.1 $\pm$ 0.6 & 2.1 $\pm$ 0.7 \\
    $Z\gamma \to ll\gamma$& 1.8 $\pm$ 0.3 & 17.6 $\pm$ 1.9 \\
    $W\gamma \to \tau\nu\gamma$& 2.3 $\pm$ 0.3 & 5.4 $\pm$ 0.6 \\
   \hline
   Total background& 39.1 $\pm$ 3.8 &89.7 $\pm$ 7.2 \\
   SM $W\gamma$ prediction& 150.9 $\pm$ 13.8 & 282.1 $\pm$ 25.4 \\
   Data& 196 & 363 \\
    \hline \hline
  \end{tabular}
  \caption{ \small
   Number of predicted and observed events with 
statistical and systematic uncertainties.
}
  \label{table:Nevents}
\end{center}
\end{table}

The sources of systematic uncertainty that affect the signal
acceptance and the background normalization include:
integrated luminosity (6.1\%), trigger efficiency (5\%),
electron identification (3\%), muon identification (3\%), 
photon identification (3\%), track veto (0.9\%), 
signal acceptance due to uncertainties on PDF (0.4\%),
predicted cross sections for $Z\gamma \to ll\gamma$ (4\%) 
and $W\gamma \to \tau\nu\gamma$ (3\%), 
and estimation of $W+$jet background (10\%).

The measured cross sections multiplied by the branching fractions for
$\sigma(p\bar{p} \to W\gamma +X \to l\nu\gamma + X)$ 
for photons
with  $E_{T}^{\gamma} >$ 15 GeV and $\Delta R_{l\gamma} >$ 0.7
are 7.9~$\pm$~0.7~(stat.)~$\pm$~0.7~(syst.)~pb for the electron channel,
and 7.4~$\pm$~0.5~(stat.)~$\pm$~0.7~(syst.)~pb for the muon channel.
The detector resolution effects that would result in 
some of the events failing the
$E_{T}^{\gamma} >$ 15 GeV and $\Delta R_{l\gamma} >$ 0.7 requirements
at the generator level but passing them at the reconstructed-object level have 
been taken into account.
Taking into account the correlation in systematic uncertainties,
the combined results yield a cross section multiplied by the branching 
fraction of 7.6~$\pm$~0.4~(stat.)~$\pm$~0.6~(syst.)~pb,
which is in good agreement with the SM expectation of $7.6 \pm 0.2$ pb. 

\begin{figure}[htbp]
 \centering
\hspace*{-9mm} \includegraphics[scale=0.41]{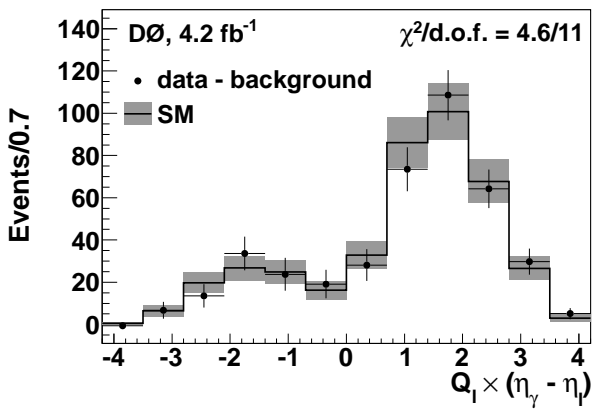}
~\\[-4mm]
   \caption{\small  
   The charge-signed photon-lepton rapidity difference 
($Q_{l} \times (\eta_{\gamma} - \eta_{l})$, where
$Q_{l}$ is the charge of the lepton)
in background-subtracted data compared to the SM expectation for
the combined electron and muon channels.
The background-subtracted data are shown as black points
with error bars representing their total uncertainties.
The SM signal prediction is given by the solid line, 
with the shaded area representing its uncertainty. 
}
  \label{figure-raz}
\end{figure}

The charge-signed photon-lepton rapidity difference
for the combination of the two channels is shown in Fig.~\ref{figure-raz}.
Because of significant charge mis-identification of EC electrons,
only events with CC electrons are used in Fig.~\ref{figure-raz}.
The background-subtracted data are in good agreement with the SM prediction,
and a $\chi^{2}$ test comparing the background-subtracted data with the SM prediction
yields 4.6 for 11 degrees of freedom.

\begin{figure}[htbp]
 \centering
\hspace*{-9mm} \includegraphics[scale=0.41]{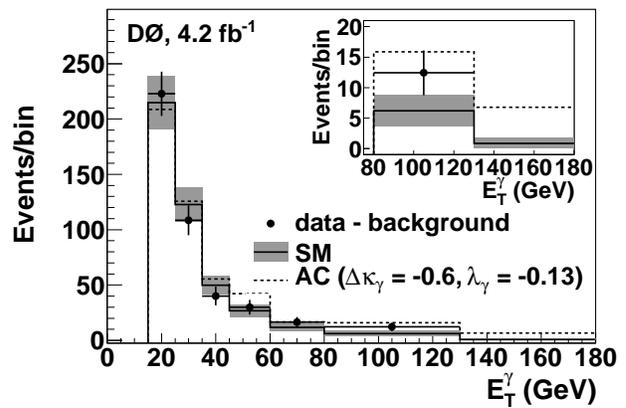}
~\\[-4mm]
   \caption{\small Photon transverse energy distributions 
for background-subtracted data compared to the expectation for the SM
and for one choice of anomalous couplings
for the combined electron and muon channels.
The background-subtracted data are shown as black points
with uncertainties representing the associated 
statistical and systematic uncertainties.
The SM prediction is given by the solid line, 
with the shaded area representing its uncertainty. 
The effect of one example of anomalous couplings 
is represented by the dashed line.
The last $E_T^{\gamma}$ bin shows the sum
of all events with $E_T^{\gamma} > 130$ GeV.
The inset shows the distributions in the last two bins of $E_T^{\gamma}$.
}
  \label{figure-gpt}
\end{figure}

\begin{figure}[htbp]
 \centering
\hspace*{-9mm} \includegraphics[scale=0.41]{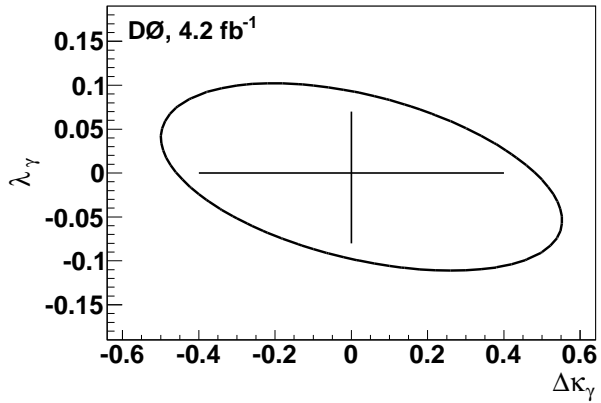}
~\\[-4mm]
   \caption{\small 
  Limits on the $WW\gamma$ coupling parameters $\Delta\kappa_{\gamma}$
and $\lambda_{\gamma}$. The ellipse represents the two-dimensional 95\% C.L.
exclusion contour. The one-dimensional 95\% C.L. limits are shown as
the vertical and horizontal lines.
}
  \label{figure-AClimits}
\end{figure}

The photon $E_{T}^{\gamma}$ distributions in Fig.~\ref{figure-gpt} 
show good agreement between data and the SM prediction.
Therefore, we use the photon $E_{T}^{\gamma}$ spectra to derive
limits on anomalous $WW\gamma$ trilinear couplings 
using a binned likelihood fit to data.
The likelihood is calculated assuming Poisson statistics for the number of events in data,
the signal, and the background. All systematic uncertainties on sources of background,
efficiencies, and luminosity are assumed to be Gaussian,
and their correlations are taken into account in the fit.
The 95\%~C.L. limits on the $WW\gamma$ coupling parameters are shown 
in Fig.~\ref{figure-AClimits}, with the contour defining the two-dimensional
exclusion limits.
The one-dimensional 95\%~C.L. limits are 
$-0.4 < \Delta\kappa_{\gamma} < 0.4$
and $-0.08 < \lambda_{\gamma} < 0.07$,
which are obtained by setting one coupling parameter to the SM value
and allowing the other to vary.

In summary, we have studied $W\gamma$ production 
using data corresponding to an integrated luminosity of 4.2~fb$^{-1}$ 
collected by the D0 detector at the Fermilab Tevatron Collider.
The cross section multiplied by the branching fraction for
the process $p\bar{p} \to W\gamma + X \to l\nu\gamma +X$ is measured to be
7.6~$\pm$~0.4~(stat.)~$\pm$~0.6~(syst.)~pb,
which is in good agreement with the SM expectation of 7.6~$\pm$~0.2~pb
for $E_{T}^{\gamma} >$ 15 GeV and $\Delta R_{l\gamma} >$ 0.7.
The distribution of the charge-signed photon-lepton rapidity difference
has a minimum near $Q_{l} \times (\eta_{\gamma} - \eta_{l})=-1/3$,
consistent with the SM prediction.
We also set the most stringent limits 
on anomalous $WW\gamma$ couplings
at a hadron collider, with
the one-dimensional parameters restricted to $-0.4< \Delta\kappa_{\gamma} < 0.4$
and $-0.08 < \lambda_{\gamma} < 0.07$, at the 95\% C.L.

%
We thank the staffs at Fermilab and collaborating institutions,
and acknowledge support from the
DOE and NSF (USA);
CEA and CNRS/IN2P3 (France);
FASI, Rosatom and RFBR (Russia);
CNPq, FAPERJ, FAPESP and FUNDUNESP (Brazil);
DAE and DST (India);
Colciencias (Colombia);
CONACyT (Mexico);
KRF and KOSEF (Korea);
CONICET and UBACyT (Argentina);
FOM (The Netherlands);
STFC and the Royal Society (United Kingdom);
MSMT and GACR (Czech Republic);
CRC Program and NSERC (Canada);
BMBF and DFG (Germany);
SFI (Ireland);
The Swedish Research Council (Sweden);
and
CAS and CNSF (China).


\begin{thebibliography}{99}
\bibitem{d0_coordinate}
Pseudorapidity is defined as $\eta=-\ln [\tan(\theta/2)]$, where
$\theta$ is the polar angle relative to the proton beam direction.
$\phi$ is defined to be the azimuthal angle in the plane transverse to 
the proton beam direction.
\end{thebibliography}
\end{document}